\documentstyle[aps,prl,preprint]{revtex}
\input epsf
\newcommand{\newc}{\newcommand}
\newc{\gsim}{\lower.7ex\hbox{$\;\stackrel{\textstyle>}{\sim}\;$}}
\newc{\lsim}{\lower.7ex\hbox{$\;\stackrel{\textstyle<}{\sim}\;$}}
\def\ms{M_{\rm SUSY}}
\def\tanb{\tan\beta}
\def\cosb{\cos\beta}
\def\sinb{\sin\beta}
\def\sb{s_{\beta}}
\def\cb{c_{\beta}}
\def\sa{s_{\alpha}}
\def\ca{c_{\alpha}}
\newc{\be}[1]
{\begin{equation} \mbox{$\label{#1}$}}
\newc{\bea}[1]
{\begin{eqnarray} \mbox{$\label{#1}$}}
\newc{\ee}{\end{equation}}
\newc{\eea}{\end{eqnarray}}
\def\lam{\lambda}
\def\Lam{\Lambda}
\def\eps{\epsilon}
\def\tw{\theta_w}
\def\sintw{\sin\tw}
\def\costw{\cos\tw}
\def\mua{\mu_{\!A}}
\def\re#1{{[\ref{#1}]}}

\begin{document}
\footnotesep=14pt
\begin{titlepage}

\begin{flushright}
\baselineskip=16pt
{\footnotesize
FERMILAB--Pub--97/102-A}\\
{\footnotesize hep-ph/9704371}\\
{\footnotesize Submitted to Phys.~Rev.~D}
\end{flushright}
\renewcommand{\thefootnote}{\fnsymbol{footnote}}
\vspace{0.15in}
\baselineskip=24pt

\begin{center}
{\Large\bf CP-Violating Solitons in the}\\ {\Large\bf Minimal Supersymmetric
Standard Model}\\
\baselineskip=16pt
\vspace{0.75cm}

{\bf Antonio Riotto\footnote{\baselineskip=16pt
PPARC Advanced Fellow,  Oxford Univ. from Sept.~1997.
{}From 1 Dec.~1997 on leave of absence at CERN, Theory Division, CH--1211
Geneva 23,
Switzerland.  Email:
{\tt riotto@fnal.gov}} and Ola T\"{o}rnkvist\footnote{\baselineskip=16pt
Address after 24 Sept.~1997: DAMTP, Univ.~of
Cambridge, Silver Street, Cambridge CB3~9EW, England.  Email: {\tt
olat@fnal.gov}}}, \\
\vspace{0.4cm}
{\em NASA/Fermilab Astrophysics Center},\\
{\em Fermi National Accelerator Laboratory},\\
{\em Batavia, Illinois~60510-0500, USA}
\vspace{0.3cm}
\vspace*{0.75cm}

{April 16, 1997}\\
\end{center}

\begin{quote}
\begin{center}
{\bf\large Abstract}
\end{center}
\vspace{0.2cm}
We study  non-topological and CP-violating static wall solutions in the
framework of the Minimal Supersymmetric Standard Model. We show that such
membranes, characterized by a non-trivial winding of the relative $U(1)$ phase
of the two Higgs fields in the direction orthogonal to the wall, exist
for small values of the mass of the CP-odd Higgs boson when
loop corrections to the Higgs potential are included. Although their
present-day
existence is excluded by experimental  bounds,
we argue
why they may have existed in the early universe with important cosmological
consequences.
\vspace*{8pt}


\end{quote}
\end{titlepage}

\newpage
\renewcommand{\thefootnote}{\arabic{footnote}}
\addtocounter{footnote}{-2}
\baselineskip=24pt
\renewcommand{\baselinestretch}{1.5}
\footnotesep=24pt

%%%%%%%%%%%%%%%%%%%%%%%%%%%%%%%%%%%%%%%%
%%%%%%%%%%%%%%%%%%%%%%%%%%%%%%%%%%%%%%%%
\subsection*{I. INTRODUCTION}
Supersymmetry provides ways to solve many of the puzzles of the Standard Model
such as the stability of the weak scale under radiative corrections as well as
the origin of the weak scale itself. Local supersymmetry provides a promising
way to include gravity within the framework of unified theories of particle
physics, eventually leading the way to a theory of everything in string
theories. Naturalness requires the masses of supersymmetric particles to be no
larger than
about
$1$ TeV, which is within the accessible range of
planned
future
 particle accelerators.
For these compelling reasons, supersymmetric extensions of the Standard Model
have been the focus of intense theoretical activity in recent years
\re{review}.

One of the basic properties of the Minimal Supersymmetric Standard Model (MSSM)
is the presence  of two Higgs  doublets which renders the scalar sector of the
low-energy theory
quite rich of consequences.  For example, a new source of CP-violation, beyond
the one contained in the CKM matrix, may appear in the Higgs sector \re{noi}
when the neutral components
of the Higgs fields acquire complex vacuum expectation values (VEV's) because
of plasma effects during the electroweak phase transition. In such a case,
particle mass matrices acquire a
nontrivial space-time dependence when bubbles of the broken
phase
nucleate and expand during a first-order electroweak phase
transition \re{reviews}.  This provides
sufficiently
fast nonequilibrium CP-violating  effects inside the wall of a
bubble of broken phase expanding in the plasma and may give
rise to a nonvanishing baryon asymmetry in the MSSM through the anomalous
$(B+L)$-violating transitions \re{sp} when particles
diffuse to the exterior of the advancing bubble [\ref{bau},\ref{toni}].

An  extended Higgs sector usually allows also for the possibility of discrete
symmetries and the presence of  associated domain walls \re{walls}.
Recently Bachas and Tomaras \re{bt} have analyzed a different
class of {\it membrane\/} defects in the generic two-Higgs-doublet model.
These electrically neutral solutions
differ from domain walls in that they interpolate between identical vacua on
either side. They are not tied to any discrete symmetry, but are instead
characterized by a
non-trivial winding,
in the direction  orthogonal to the wall,
 of the relative $U(1)$ phase $\theta$ of the two Higgs fields. Such solutions
arise
because of the presence of a term proportional to $\cos\theta$  in the bilinear
part of the Higgs
potential. In the $\sigma$-model limit, where the neutral Higgs components are
held fixed at their
VEV's, the membrane indeed coincides with the kink solution of the sine-Gordon
model.

The typical thickness of the membranes is $M_A^{-1}$, the inverse mass of the
CP-odd
Higgs scalar $A^0$, and whereas they are not topologically stable, they may
have
a finite lifetime. Although  the analysis performed in \re{bt} was restricted
to the case $\tan\beta=1$,
$\tan\beta=v_2/v_1$ being the ratio  of  the VEV's $v_2$ and $v_1$ of the two
neutral Higgs components, it is important to mention that, as a general
property, these CP-violating
membranes seem to exist only when the
CP-odd scalar $A^0$ is the lightest neutral eigenstate in the scalar sector.
Therefore, it was concluded in Ref.~\re{bt} that the MSSM lies outside the
region of existence of membranes since,
 at  tree level, $A^0$ is generally more massive than the lightest CP-even
scalar
$h$.
However, as we shall demonstrate,  supersymmetry constraints
on  the Higgs sector  are so restrictive that  CP-violating membranes do not
exist at all for any values of $M_A$ and  $\tan\beta$ at  tree level. This
means that no conclusion may be drawn {\it a priori\/} about the existence of
membranes in the framework of MSSM from considerations about the scalar
spectrum unless one relaxes the tree-level conditions in the Higgs scalar
sector by including loop corrections to the Higgs potential.

The purpose of this paper is to study  the
CP-violating membrane solutions within the  MSSM and to show that  these
solutions  exist (only) when loop corrections to the Higgs potential are taken
into account and that, as was speculated in Ref.~\re{bt}, they exist only when
the
CP-odd scalar $A^0$ is lighter than the lightest CP-even scalar $h$. This is
made
possible because
loop corrections coming from the top-stop sector  considerably modify the
hierarchy in the scalar spectrum at  tree level \re{quiros} and allow the
relation $M_A<M_h$.

The paper is organized as follows. Section II contains the description of the
model and the relevant equations. In Section III we analytically investigate
the existence of solutions.
Section IV is devoted to the presentation of numerical solutions and results.
Finally,
Section V contains our conclusions and a discussion about possible cosmological
implications of the CP-violating membranes.

\subsection*{II. THE MODEL}
We denote the two Higgs doublets of the model by
\be{doublets}
H_1\equiv \left(\begin{array}{c} H_1^1\\*[-1ex]H_1^2\end{array} \right)
\equiv \left(\begin{array}{c} \phi_1\\*[-1ex]\phi_1^-\end{array} \right);
\quad H_2\equiv \left(\begin{array}{c} H_2^1\\*[-1ex]H_2^2\end{array} \right)
\equiv \left(\begin{array}{c} \phi_2^+\\*[-1ex]\phi_2\end{array} \right)
\ee
with hypercharge  $y=-1, 1$ respectively. The components
$\phi_1$ and $\phi_2$ are electrically neutral.
The Lagrangian is
\be{lag}
{\cal L}=| D_\mu H_1|^2+ |D_\mu H_2|^2  -
V(H_1,H_2)\  - \frac{1}{4} W^a_{\mu\nu} W^{a\mu\nu} -
\frac{1}{4} Y_{\mu\nu} Y^{\mu\nu},
\ee
where $D_\mu \equiv \partial_\mu +\frac{i}{2}g W_\mu^a\tau^a +
\frac{i}{2}g'yY_\mu$ and the most general gauge-invariant potential
$V(H_1,H_2)$ is given
by
\bea{pot}
V(H_1,H_2) &=& m_1^{\,2} |H_1|^2 + m_2^{\,2} |H_2|^2 - m_3^{\,2} [
(H_1 H_2)  +
 {\rm h.c.} ] \nonumber\\* &+ &\lam_1 |H_1|^4 + \lam_2 |H_2|^4
+ \lam_3 |H_1|^2 |H_2|^2 + \lam_4 |(H_1 H_2)|^2 \nonumber\\*&+&
\left[ \lam_5 (H_1 H_2)^2 + \lam_6  |H_1|^2 (H_1 H_2) +
\lam_7 |H_2|^2(H_1 H_2) +  {\rm h.c.}\right]
\eea
with $(H_1 H_2)\equiv \eps_{ij} H_1^i H_2^j$ and
$|H_I|^2 \equiv H_I^{\dag} H_I$, ${\scriptstyle I}=1,2$.  In eq.~(\ref{lag})
we have
$W^a_{\mu\nu} = \partial_\mu W^a_\nu - \partial_\nu W^a_\mu - g
\eps^{abc}W^b_\mu W^c_\nu$  and
$Y_{\mu\nu} = \partial_\mu Y_\nu - \partial_\nu Y_\mu$. The physical $Z^0$ and
photon fields are given by $Z_\mu = W^3_\mu \costw - Y_\mu \sintw$ and
$A_\mu = W^3_\mu \sintw + Y_\mu \costw$, where the weak mixing angle
$\tw$ satisfies $\tan\tw = g'/g$.

The minimum of the potential (\ref{pot}), for values of the parameters
that yield  positive squared masses of the physical Higgs bosons, is
given by the vacuum
\be{minpot}
\langle H_1\rangle = \frac{1}{\sqrt{2}}\left(\begin{array}{c} v_1\\*[-1ex]
0\end{array} \right);
\quad \langle H_2\rangle\equiv \frac{1}{\sqrt{2}}
\left(\begin{array}{c} 0\\*[-1ex]  v_2 e^{i\delta}\end{array} \right)\ ,
\ee
where $v_1^2 + v_2^2 = v^2$ and $v=246.2$ GeV is fixed by the mass of the $W$
boson, $M_W^2 = g^2 v^2/4$.  The constant phase
$\delta$, when it is not a multiple of $\pi$,  provides a spatially uniform
source of CP violation.
On the other hand, the couplings  $\lam_5$, $\lam_6$, and  $\lam_7$ are zero at
tree
level and receive
small loop corrections that  may be neglected in the present context  without
affecting
any of the conclusions.  In such a case $\delta=0$, and there is no
``background'' CP violation in the model. CP violation will occur only inside
the
membranes where there will be an additional, space-dependent relative phase
$\theta$.

We shall consider  static solutions to the field equations in
which
only the neutral fields $\phi_1$, $\phi_2$ and $Z_\mu$ participate. It can be
easily verified that the system of field equations for $\phi_1^-$,  $\phi_2^+$,
$W^1_\mu$, $W^2_\mu$ and $A_\mu$ is homogeneous, and thus permits solutions
where these fields are identically zero.
The resulting Lagrangian is
\be{lagn}
{\cal L}^0=\left| (\partial_\mu  + \frac{ig}{2\costw} Z_\mu) \phi_1\right|^2+
\left| (\partial_\mu  - \frac{ig}{2\costw} Z_\mu) \phi_2\right|^2-
V^0(\phi_1,\phi_2)\   -
\frac{1}{4} Z_{\mu\nu} Z^{\mu\nu}\ ,
\ee
where $Z_{\mu\nu}\equiv \partial_\mu Z_\nu - \partial_\nu Z_\mu$ and
\bea{potn}
V^0(\phi_1,\phi_2) &=&
m_1^{\,2} |\phi_1|^2 + m_2^{\,2} |\phi_2|^2 - m_3^{\,2} [  \phi_1
\phi_2  +
 {\rm c.c.} ] \nonumber\\* &+ &\lam_1 |\phi_1|^4 + \lam_2 |\phi_2|^4
+ (\lam_3 + \lam_4)|\phi_1|^2 |\phi_2|^2\ .
\eea

  The couplings $\lam_1,\ \lam_2,\ \lam_3$ and $\lam_4$ are
determined by supersymmetry.  By minimizing the potential
(\ref{potn}) the quantities $m_1^{\,2}$,  $m_2^{\,2}$
and $m_3^{\,2}$ may be reexpressed in terms of the electroweak scale $v$,
the ratio of Higgs expectation values $\tan\beta\equiv v_2/v_1$ and the
mass $M_A$ of the neutral CP-odd scalar $A^0$. We have
\bea{m2par}
m_1^{\,2} &=& \left(M_A^{\,2}-\frac{1}{2}(\lam_3+\lam_4)v^2 \right)\sin^2\beta
 - \lam_1 v^2\cos^2\beta \ , \nonumber\\*
m_2^{\,2} &=& \left(M_A^{\,2}-\frac{1}{2}(\lam_3+\lam_4)v^2 \right)\cos^2\beta
 - \lam_2 v^2\sin^2\beta \ , \nonumber\\*
m_3^{\,2} &=& \frac{1}{2}M_A^{\,2}\sin 2\beta\ .
\eea
The parameters $\beta$ and $M_A$ therefore
completely parametrize the model at tree level.
When one-loop corrections are included, at least one more parameter is needed.

Variation of the action $\int\!dx\, {\cal L}^0$ with respect to $Z_\mu$,
$\phi_1^{\ast}$ and $\phi_2^{\ast}$
gives
\bea{eleq1}
\partial_\mu  Z^{\mu\nu} +
\frac{g}{2\costw} [
 - i \phi_1^{\ast} (\partial_\nu+iz_\nu)\phi_1 +
   i \phi_2^{\ast} (\partial_\nu-iz_\nu)\phi_2 + {\rm c.c.}]=0\ ,\\*
\label{eleq2}
(\partial_\mu+i z_\mu)^2 \phi_1+m_1^{\,2} \phi_1-m_3^{\,2}
\phi_2^{\ast}
+ 2\lam_1|\phi_1|^2\phi_1 + (\lam_3+\lam_4) |\phi_2|^2\phi_1 = 0\ ,\\*
\label{eleq3}
(\partial_\mu-i z_\mu)^2 \phi_2+m_2^{\,2} \phi_2-m_3^{\,2}
\phi_1^{\ast}
+ 2\lam_2|\phi_2|^2\phi_2 + (\lam_3+\lam_4) |\phi_1|^2\phi_2 = 0\ ,
\eea
where $z_\mu \equiv g Z_\mu/(2 \costw)$.

We now turn to the ansatz for the static membrane solution with unit
winding number. We consider the simplified case of an infinitely large, flat
membrane.
The fields then depend only on the coordinate $x$ perpendicular to the
membrane:
\be{ansatz}
\phi_1(x) = \frac{v_1}{\sqrt{2}} f_1(x) e^{i\theta(x)}\ , \quad
\phi_2(x) = \frac{v_2}{\sqrt{2}} f_2(x) \ , \quad
Z_x = \frac{2\costw}{g}\, z(x)\ ,
\ee
where $f_1(\pm\infty)=f_2(\pm\infty)=1$, $\theta(-\infty)=0$ and
$\theta(\infty)=2\pi$. In order to fix the position of the membrane at $x=0$
(for example),
it is
necessary to consider the symmetries of the differential equations when
$x\to -x$, impose instead boundary conditions
$f_1^{\,'}(0)=f_2^{\,'}(0)=0$,
$\theta(0)=\pi$ and solve the problem on the positive semi-infinite interval.

In one dimension, the field tensor $Z_{\mu\nu}$ is identically zero, and so
eq.~(\ref{eleq1}) turns into a constraint that relates the unphysical
(pure-gauge) field $Z_x$ to the gradient of the phase $\theta$:
\be{zeq}
z = - \frac{f_1^{\,2}\cos^2\beta}{f_1^{\,2}\cos^2\beta+f_2^{\,2}\sin^2\beta}
\,
\frac{d\theta}{d x}\ .
\ee
The
gauge choice $z=0$, used in Ref.~\re{bt}, is simple
only for $\tanb=1$.

By inserting the functional forms (\ref{ansatz}) into eqs.~(\ref{eleq2}) and
(\ref{eleq3}), making use of eq.~(\ref{zeq}),
and extracting the real and imaginary parts, the system of differential
equations can
be written
\bea{eq1}
f_1^{\,''}&=&\mua^{\,2} \sb^{\,4} \frac{\sigma^2}{f_1^{\,3}} +
\frac{m_1^{\,2}}{M_Z^{\,2}} f_1 + \cb^{\,2}\Lam_1 f_1^{\,3} +
\sb^{\,2}\Lam_{34} f_2^{\,2} f_1 - \mua^{\,2} \sb^{\,2} f_2 \cos\theta\ ,\\*
\label{eq2}
f_2^{\,''}&=&\mua^{\,2} \cb^{\,4} \frac{\sigma^2}{f_2^{\,3}} +
\frac{m_2^{\,2}}{M_Z^{\,2}} f_2 + \sb^{\,2}\Lam_2 f_2^{\,3} +
\cb^{\,2}\Lam_{34} f_1^{\,2} f_2 - \mua^{\,2} \cb^{\,2} f_1 \cos\theta\ ,\\*
\label{eq3}
\theta^{\,'}&=&\mua\left( \frac{\sb^{\,2}}{f_1^{\,2}} +
\frac{\cb^{\,2}}{f_2^{\,2}}\right)\sigma\ ,\\*
\label{eq4}
\sigma^{\,'}&=&\mua f_1 f_2 \sin\theta \ .
\eea
Here a prime ($'$) indicates a derivative with respect to the dimensionless
coordinate
$y\equiv M_Z x$, and $\sigma$ is an auxiliary field defined by
eq.~(\ref{eq3}).
The various constants are defined as follows:
\be{cdef}
\mua= \frac{M_A}{M_Z};\   \sb=\sinb;\  \cb=\cosb;
\ \Lam_1= \frac{\lam_1 v^2}{M_Z^{\,2}};\
\Lam_2= \frac{\lam_2 v^2}{M_Z^{\,2}};\
\Lam_{34}= \frac{\lam_3+\lam_4}{2 M_Z^{\,2}}\ .
\ee
The field variables $f_1$, $f_2$, $\theta$ and $\sigma$ have been scaled in
such
a way that their typical order of magnitude is unity.

Note here that, in the limit where $f_1\equiv f_2\equiv 1$, eqs.~(\ref{eq3})
and
(\ref{eq4})  reduce to the equation for the  sine-Gordon
kink (or circular pendulum), $\theta^{\,''} - \mua^2 \sin\theta=0$, with
analytic
solution $\theta = 4\tan^{-1}[\exp(\mua y)] =4\tan^{-1}[\exp(M_A x)]$
corresponding
to a membrane of characteristic thickness $M_A^{-1}$.

Because  eqs.~(\ref{eleq2}) and
(\ref{eleq3}) constitute four {\em real\/} second-order differential
equations,
there is one second-order equation missing.
It corresponds to the CP-odd Goldstone mode, and is simply
\be{goldst}
\frac{d}{dx}\left[z (\cb^{\,2}f_1^{\,2} + \sb^{\,2}f_2^{\,2}) +  \cb^{\,2}
f_1^{\,2}
\frac{d\theta}{dx}
\right]=0\ .
\ee
This is merely an integrability condition consistent with eq.~(\ref{zeq})\ .

In the following, we shall assume that $M_W\ll \ms\lsim {\cal O}({\rm few})$
TeV,
where $\ms$ is the characteristic supersymmetry particle mass scale.
Higher values of $\ms$ would conflict with naturalness. We neglect mass mixing
in the stop sector, and
assume that all
supersymmetric
particle masses are of order $\ms$. In this approximation, and for
$M_A\lsim \ms$, accurate analytical low-energy
approximations to the one-loop radiative
corrections to the coupling constants $\lam_i$, $i=1,\ldots,7$, have been
derived by Carena et al.~\re{Carena} in terms of the parameter $t$, where
\be{tdef}
t=\ln \frac{\ms^{\,2}}{M_t^{\,2}}
\ee
and $M_t$ is the top-quark mass.
Here, we shall make also the assumption that the supersymmetric
Higgs mass $\mu$  as well as the soft trilinear
supersymmetry-breaking parameters $A_t$, $A_b$ and $A_{tb}$ are small
compared to $\ms$.  This justifies our setting
$\lam_5=\lam_6=\lam_7=0$. In addition, we may safely neglect the Yukawa
couplings
of all flavors except the top (stop). This coupling is given by $h_t^{\,2} =
h^2/\sin^2\beta$ where $h^2 \equiv 2 (M_t/v)^2\approx 1$. From
Ref.~\re{Carena}
we then obtain
\bea{lamcorr1}
\Lam_1& = & \frac{1}{2}\ ,\\*
\label{lamcorr2}
\Lam_2&=& \frac{1}{2}\left(1 - \frac{3}{8\pi^2}\frac{h^2}{\sb^{\,2}}\,
t\right)
+ \frac{3}{16\pi^2}\frac{h^4}{\sb^{\,4}}\left(\frac{v}{M_Z}\right)^2
\!\left[t + \frac{1}{16\pi^2}\left(\frac{3 h^2}{2\sb^{\,2}} - 8
g_{\rm s}^{\,2}\right) t^2\right]\ ,\\*
\label{lamcorr34}
\Lam_{34}&=&-\frac{1}{2} \left(1 -
\frac{3}{16\pi^2}\frac{h^2}{\sb^{\,2}} \,t\right)\ ,\\*
\label{lamcorr4}
\lam_4 &=& g^2 \Lam_{34}\ ,
\eea
where $g_{\rm s}$ is the strong coupling constant. Here the couplings $g$ and
$g_{\rm s}$
are meant to be computed at  the scale $M_t$.

Using eqs.~(\ref{m2par}), (\ref{cdef}), (\ref{lamcorr1}-\ref{lamcorr4}), all
quantities can
now be expressed in terms of the three parameters $\mua$, $\tanb$, and $t$.
For
example, the mass matrix of the physical neutral CP-even Higgs bosons
is
\be{mhmatrix}
{\cal M}^2 = M_Z^{\,2} \left(\begin{array}{cc}
\mu_{11}^{\,2}&\mu_{12}^{\,2}\\*
\mu_{12}^{\,2} &\mu_{22}^{\,2}
\end{array}\right)\ ,
\ee
where $\mu_{11}^{\,2} = 2 \Lam_1 c_\beta^{\,2} + \mua^{\,2}s_\beta^{\,2}$,
$\mu_{22}^{\,2} = 2 \Lam_2 s_\beta^{\,2} + \mua^{\,2}c_\beta^{\,2}$,
$\mu_{12}^{\,2}  = (2\Lam_{34} - \mua^{\,2})\sb\cb$.
  The mass eigenstates  are
\bea{Hhdef}
H&=& \cos\alpha\cos\beta\,(f_1-1) + \sin\alpha\sin\beta \,(f_2-1)\\*
h&=& -\sin\alpha\cos\beta\, (f_1-1) + \cos\alpha\sin\beta\, (f_2-1)\ ,
\eea
where the Higgs mixing angle $\alpha$ satisfies $-\pi/2\leq\alpha\leq 0$ and
is given by
\be{alphadef}
\sin 2\alpha =
{{2 \mu_{12}^2
}\over{\sqrt{(\mu_{11}^2 - \mu_{22}^2)^2 + 4 \mu_{12}^4}
}}\  ;\quad\quad
\cos 2\alpha =
{{\mu_{11}^2 -\mu_{22}^2
}\over{\sqrt{(\mu_{11}^2 - \mu_{22}^2)^2 + 4 \mu_{12}^4}
}}\  .
\ee
The  mass eigenvalues for $H$ and $h$
are
\be{mhiggs}
M_{H,h}^{\,2} = \frac{1}{2}M_Z^{\,2} \left[\mu_{11}^{\,2} + \mu_{22}^{\,2}
\pm \sqrt{ (\mu_{11}^{\,2} - \mu_{22}^{\,2})^2 + 4 \mu_{12}^{\,4} }\, \right]\
{}.
\ee
At  tree level the mass of the lighter Higgs boson $h$ is bounded to be smaller
than both $M_Z|\cos2\beta|$ and $M_A$. This conclusion is modified by radiative
corrections which raise the upper limit on the lightest CP-even Higgs mass to
values near 150 GeV.

The mass of the charged Higgs particles $H^{\pm}$ is
\be{mhpm}
M_{H^{\pm}}^{\,2} = M_A^{\,2} - \frac{1}{2} \lam_4 v^2 =
M_A^{\,2} + M_W^{\,2}\left(1 -
\frac{3}{16\pi^2}\frac{h^2}{\sb^{\,2}}\, t\right)\ .
\ee
For high values of $\ms$, this squared mass becomes negative for
 low values of  $\tanb$,
indicating that the potential for such parameter values no longer has a minimum
of the
type (\ref{minpot}) with $v_1,\,v_2\neq 0$.
 All other particle squared masses remain
non-negative.

Using the above expressions one can now describe the asymptotic behavior of the
different fields. Let us define $\mu_h = M_h/M_Z$, $\mu_H = M_H/M_Z$.
The leading terms of the equations (\ref{eq1})--(\ref{eq4}) in the
asymptotic regime of large $y$ are
\bea{asymp}
\theta^{\,''} + \mua^2 \,(2\pi-\theta) &=& 0,\\*
h^{\,''} - \mu_h^2 h &=& c_\theta^2 a_A e^{-2\mua y} + a_H H^2\\*
H^{\,''} - \mu_H^2 H &=& c_\theta^2 b_A e^{-2\mua y} + b_h h^2
\eea
where $\theta \sim 2\pi - c_\theta e^{-\mua y}$ is the  solution of
eq.~(\ref{asymp})  and
\bea{abdef}
a_A &=& \frac{1}{2}\mua^2 w \sb\cb
\left[ \ca\cb (1 + 2 \cb^{\,2}) - \sa\sb (1+2\sb^{\,2})\right]\nonumber\\*
b_A &=& \frac{1}{2}\mua^2 w \sb\cb
\left[ \sa\cb (1 + 2 \cb^{\,2}) + \ca\sb (1+2\sb^{\,2})\right]\nonumber\\*
a_H & = & -\sa\cb[3(\Lam_1-\Lam_{34})\ca^{\,2} + \Lam_{34}] +
\ca\sb[3(\Lam_2-\Lam_{34})\sa^{\,2} + \Lam_{34}]\nonumber\\*
b_h & = & +\ca\cb[3(\Lam_1-\Lam_{34})\sa^{\,2} + \Lam_{34}] +
\sa\sb[3(\Lam_2-\Lam_{34})\ca^{\,2} + \Lam_{34}]
\eea
 with $\ca=\cos\alpha$, $\sa=\sin\alpha$, and $ w=1$.

The characteristic exponential for $h$ and $H$ is determined by the
particular solutions of the inhomogeneous
equations as well as the homogeneous
solutions. Assuming a characteristic asymptotic behavior $h\sim e^{-\bar{\mu}_h
y}$,
$H \sim e^{-\bar{\mu}_H y}$, we get
$\bar{\mu}_h = \min (\mu_h,2\mua,2\bar{\mu}_H)$ and
$\bar{\mu}_H = \min (\mu_H,2\mua,2\bar{\mu}_h)$. Since $\mu_h\leq \mu_H$ we
obtain
\bea{muineq}
\bar{\mu}_h &= &\min (\mu_h,2\mua)\ , \nonumber\\*
\bar{\mu}_H &=& \min (\mu_H,2\mua,2\mu_h)\  .
\eea
It can then be shown that
the next-to-leading asymptotic terms in $\theta$, $H$, and $h$ are suppressed
by
at least a
factor $\exp(-\bar{\mu}_h y)$.
\newpage
\subsection*{III. ANALYTICAL INVESTIGATION OF THE EXISTENCE
OF SOLUTIONS}

The equations (\ref{eq3}) and (\ref{eq4}) have bounded solutions, satisfying
$\theta(-\infty)=0$ and $\theta(\infty)=2\pi$, for any positive functions
$f_1$ and $f_2$ with $f_1(\pm\infty)=f_2(\pm\infty)=1$. We  therefore
focus on the question of existence of solutions to eqs.~(\ref{eq1}) and
(\ref{eq2}).

The energy density of the membrane solution contains the terms
$M_Z^{\,2} v^2{\theta^{\,'}}^2 [\cb^{-2} f_1^{-2} +
\sb^{-2}f_2^{-2}{]}^{-1}  /2$ and
$-M_A^{\,2} v^2 \sb^{\,2}\cb^{\,2}  f_1 f_2 \cos\theta$ whereby the phase
field
$\theta$ interacts with the magnitudes $f_1$ and $f_2$. Because
$\theta(0)=\pi$
and $\theta^{\,'}$ is expected to peak at $x=0$,
a static solution corresponding to a minimum of the energy functional must
reduce the contribution to the energy from these two positive
terms
by forcing $f_1(0)<1$ and $f_2(0)<1$.
Then, since $f_{1,2}\to 1$ asymptotically as $x\to \pm\infty$, we must have
positive curvature at the origin, $f_{1,2}^{\,''}(0) >0$, as well as
negative curvature  $f_{1,2}^{\,''}(x) <0$ for large $|x|$.

The first condition is very easy to achieve through the positive definite
terms $\sigma^2/f_{1,2}^{\,3}$
in eqs.~(\ref{eq1}) and (\ref{eq2}), either by a high value of
$\theta^{\,'}(0)$ or low values
of  $f_{1,2}(0)$,  and can be shown to impose no appreciable restrictions on
the parameters.

In order to examine the possibility of negative curvature, we use
eq.~(\ref{m2par}) to
rewrite
eqs.~(\ref{eq1}) and (\ref{eq2}) in the following form:
\bea{neweq1}
f_1^{\,''}&=&\mua^{\,2} \sb^{\,4} \frac{\sigma^2}{f_1^{\,3}} +
\cb^{\,2}\Lam_1(f_1^{\,2} - 1) f_1 + \sb^{\,2}\Lam_{34} (f_2^{\,2} - 1) f_1
+  \mua^{\,2} \sb^{\,2} (f_1 -f_2 \cos\theta)\\*
\label{neweq2}
f_2^{\,''}&=&\mua^{\,2} \cb^{\,4} \frac{\sigma^2}{f_2^{\,3}} +
\sb^{\,2}\Lam_2(f_2^{\,2} - 1) f_2 + \cb^{\,2}\Lam_{34} (f_1^{\,2} - 1) f_2
+  \mua^{\,2} \cb^{\,2} (f_2 -f_1 \cos\theta)
\eea
We consider these two equations in the region of large $|x|$, where
$1-\eps_1 < f_1\approx f_2< 1$ and $1-\eps_2 < \cos\theta < 1$ for small
positive numbers $\eps_1$ and $\eps_2$.

We first notice that a low value of $\mua$ will prevent the positive
definite $\sigma^2$ term in both equations from becoming too large. This
condition
also reduces the influence of the last term.

At tree level $(t=0)$ we have $\Lam_1=\Lam_2=1/2$ and $\Lam_{34}=-1/2$.
Then the equations have no solution for any value of the parameters
$\mua$ and $\tanb$. In order to show this, consider first the case of
$\tanb=1$.
Then from symmetry we have  $f_1=f_2=f$ which should satisfy
\be{tanb1eq}
f^{\,''} = \left[\frac{1}{4}\mua^{\,2}\frac{\sigma^2}{f^4} + \frac{1}{2}
\mua^{\,2} (1-\cos\theta)\right] f\ .
\ee
The right-hand side of this equation is positive definite\footnote{The
magnitude $f$ is
by definition non-negative. If it should ever reach zero, the phase $\theta$
would be undefined.} which makes it impossible
to have a solution.

Consider next $\tanb\neq 1$ at tree level. In eq.~(\ref{neweq1}) we have
$\cb^{\,2}\Lam_1(f_1^{\,2} - 1) f_1 <0$ and $\sb^{\,2}\Lam_{34} (f_2^{\,2} - 1)
f_1>0$,
where the magnitude of the two terms is comparable for $\tanb=1$.
We could therefore make the negative term dominate (also over
the $\sigma^2$ term) by choosing $\tanb$ sufficiently small.
But this leads to trouble in  eq.~(\ref{neweq2}), where
$\sb^{\,2}\Lam_2(f_2^{\,2} - 1) f_2 <0$ and $\cb^{\,2}\Lam_{34} (f_1^{\,2} - 1)
f_2>0$.
And vice versa.

When we include the radiative corrections, however, there is a way out. We see
this by recognizing that $\Lambda_2$ gets the largest
 contribution from radiative corrections. As a result, the constant $\Lam_2$ is
a positive,
monotonically increasing function of $t$
for realistic values of  $t$ and $\tanb$. Negative curvatures can  be achieved
for both $f_1$ and $f_2$ at large $|x|$ by choosing  a low value of $\tanb$
that makes
the negative term $\cb^{\,2}\Lam_1(f_1^{\,2} - 1) f_1$ dominate in
eq.~(\ref{neweq1}),
while choosing a large value of  $t$ so as to make
the negative term $\sb^{\,2}\Lam_2(f_2^{\,2} - 1) f_2$ dominate in
eq.~(\ref{neweq2}).
Note that the low value of $\tanb$ also helps create a large value of
$\Lam_2$.

Our  conclusion is therefore that   necessary conditions for  the
existence of solutions
of the field equations are  low values of $\mua$ and $\tanb$, as well as a
sufficiently
high value of $t$ (i.e.\ $\ms$).  At tree level, no solutions exist.

\subsection*{IV. NUMERICAL SOLUTIONS}

We have solved the field equations (\ref{eq1})--(\ref{eq4})
numerically by the method of relaxation \re{numrec} of the
corresponding system of finite difference equations, using a
dynamically adaptive grid in the independent variable $y=M_Z x$
\re{Eggleton}. We have found this method particularly
reliable and worth the extra programming effort,
as it does not attempt to converge to false solutions in regions
of parameter space where none exist. For convergence
the results of two successive iterations  were required to differ by
less than $5\cdot10^{-6}$ in each field. The functions were
taken to
satisfy the boundary conditions $\theta(0)=\pi$, $f_1^{\,'}(0)=
f_2^{\,'}(0)=0$ and $\theta(R)=2\pi$, $f_1(R)=f_2(R)=1$, where $R$ is
a number chosen large enough that the inflicted relative error in each boundary
conditon is smaller than $10^{-5}$.

In eqs.~(\ref{eq1}), (\ref{eq2}) the quantity $\sigma^2$ was replaced with
$w \sigma^2$, and  $1 + w(\cos\theta -1)$ was substituted for $\cos\theta$,
where $w$ takes values in $[0,1]$. For
$w=0$ the system has the sine-Gordon kink solution $\theta=
4\tan^{-1}[\exp(\mua y)]$, $f_1=f_2\equiv 1$, while for
$w=1$ it is the true system for which a solution is sought.
The solution was obtained by taking small steps in the parameter
$w$, using each previously obtained solution as a new initial guess.
In this method lies the assumption that any solution is continuously
connected to the sine-Gordon kink. Because the field $\theta$ in both
cases satisfies boundary conditions which enforce the presence of
a kink, such an assumption is most natural. In all the solutions found,
the field $\theta$ indeed deviates very little from the sine-Gordon
kink solution.

\begin{figure}
\epsfbox{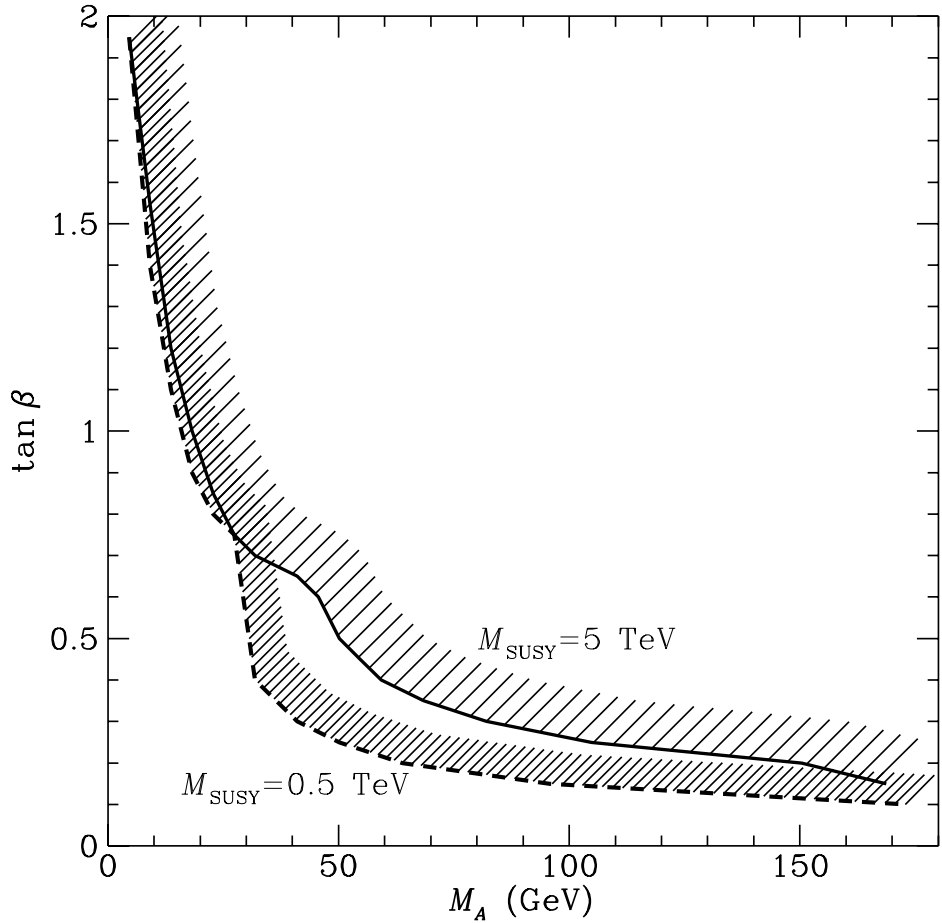}
\caption{
Region of parameter space
$(M_A,\tanb)$ where membrane solutions exist, for two different values of
the supersymmetry-breaking mass $\ms$. Solutions exist below and to the
left of  the curves.
}
\label{figure1}
\end{figure}

Solutions were sought for parameters in the ranges
$0\leq \mua\equiv M_A/M_Z\leq 10$,
$0.1 < \tanb \leq 10$, and $0\leq t \equiv\ln(\ms^{\,2}/M_t^{\,2})\leq 10$. In
agreement with the qualitative discussion of the previous section,
we found no solutions for $t=0$ (tree level). For realistic values
of $t$, corresponding to values of $\ms$ between $500$ GeV and
$5$ TeV, solutions were found for
$M_A\lsim 50$ GeV and for $\tanb\lsim 0.5$. The region of parameter space
($M_A$, $\tanb$)  where solutions exist
is shown in Fig.~1. Fig.~2 depicts a typical solution.

\begin{figure}
\epsfbox{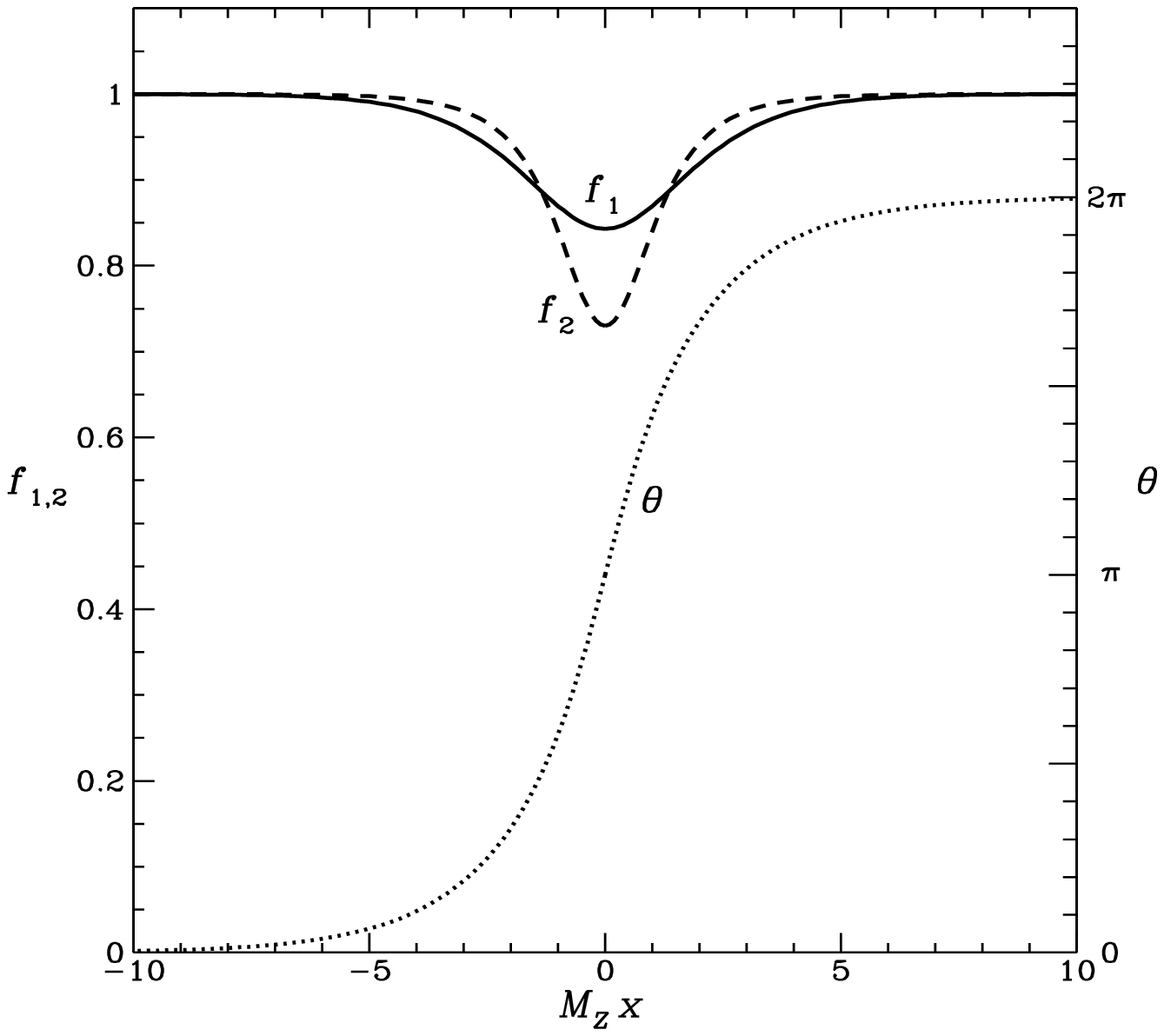}
\caption{The membrane solution for $M_A=50$ GeV, $\tanb=0.5$ and
$\ms=5$ TeV.}
\label{figure2}
\end{figure}

We used $M_t=175$ GeV, $v=246.2$ GeV, $M_Z=91.2$ GeV,
$M_W=80.2$ GeV, and
$ g_{\rm s}^2/4\pi \equiv\alpha_{\rm s} =\alpha_{\rm s}(M_t)=0.107$.

The solutions were unchanged as the number of grid points was
doubled, and it was verified that they obey integral sum rules akin to the
virial theorem.  An independent
run with $w$ fixed at $1$, taking as initial guesses the true
solutions for adjacent values of $(\mua,\tanb,t)$
 rather than the sine-Gordon kink, gave the same region of
existence of solutions.
This region  is also quite insensitive to changes in the
value of the top-quark mass in the range $160$ GeV$\leq M_t \leq 190$ GeV.

\vspace*{-5mm}
\subsection*{V. CONCLUSIONS AND OUTLOOK}

In this paper we have presented the results of a detailed investigation of
non-topological and CP-violating static wall solutions in the
framework of the Minimal Supersymmetric Standard Model. We have shown  that
membranes, characterized by a non-trivial winding of the relative $U(1)$ phase
of the two Higgs fields in the direction orthogonal to the wall, do not exist
when the Higgs potential is computed at tree level, but appear  when quantum
loop corrections to the Higgs potential are included.

Our results demonstrate that CP-violating membranes exist only for small values
of
the mass of the CP-odd Higgs boson. This does not come as a surprise. Indeed,
it was shown on general grounds by Georgi and Pais \re{gp} that gauge theories
with
perturbative spontaneous symmetry breaking may exhibit CP violation solely by
the
structure of quantum corrections to the tree-level potential. This may occur if
and only if there exist light pseudoscalars at the one-loop level. Even though
the
Georgi-Pais theorem was proved only for spatially uniform CP-violating ground
states,
we conjecture that a similar conclusion may be attained for CP-violating
solitons whose
existence is due to quantum effects alone. In our case, the light pseudoscalar
should be
identified with the CP-odd Higgs boson.

We have presented our
analysis in
terms of the model parameters  $\tan\beta$ and $M_A$; see Fig.~1. Let us now
compare our results to the present experimental and theoretical bounds in the
$(\tan\beta,M_A)$ plane.

The value of $\tan\beta$ may be theoretically bounded from below by invoking
some
ideas from grand unified scenarios.
Indeed, if one assumes the perturbative validity of the MSSM up to a
scale of $\sim 10^{16}$ GeV (the so-called ``desert'' hypothesis), the
low-energy value of the top
Yukawa coupling $h_t$ no longer depends upon its `initial' value at high scale.
This is known as the quasi-infrared fixed-point solution and gives a
theoretical prediction of  the physical top-quark mass $M_t$ that, when
combined  with the
experimental bound $M_t=(175\pm 6)$ GeV, leads to the bound\footnote{\samepage
This bound can be  lowered slightly, to
$\tan\beta>1.1$,
in gauge-mediated SUSY-breaking models, due to the presence of additional
colored matter fields at the intermediate scale $M\sim 10^{5}-10^{7}$ GeV.}
$\tan\beta>1.4$ .
The most recent experimental bound on $M_A$ has been given by the ALEPH
Collaboration from the LEP run at 172 GeV \re{aleph}. Combining results from
the   channel productions $e^{+} e^{-}\rightarrow hA,hZ$ exclude a CP-odd Higgs
boson lighter than about 62.5 GeV for $\tan\beta>1$. Comparing these bounds
with the existence curves in Fig.~1, we may conclude that CP-violating
membranes do not exist in the allowed region of  parameter space
$(M_A,\tan\beta)$.

Despite  the fact  that the existence of  these objects
is experimentally ruled out \mbox{{\em today\/}}, we argue here that they may
have existed and played a significant role during the electroweak phase
transition.

The basic parameter which  controls the existence of membranes is the squared
mass $m_3^{\,2}$ which multiplies the operator $H_1H_2$ in the Higgs potential
(\ref{pot}). It is connected to the physical CP-odd Higgs boson mass by the
relation $m_3^{\,2}=M_A^{\,2}\sin 2\beta/2$. From our results we may conclude
that
CP-violating membranes exist (for $\tan\beta>1$) only if $m_3^{\,2}$ is very
small,
in contradiction with experimental bounds. However, plasma  corrections coming
from the thermal bath that constitutes the early Universe  at  temperature $T$
may drastically alter this conclusion. As a matter of fact, the
zero-temperature parameter $m_3^{\,2}$ receives a large temperature-dependent
correction
$\Delta m_3^{\,2}(T)$ from the interactions of the Higgs fields  with stops,
charginos and neutralinos,
which populate the plasma for temperatures larger than   about 100 GeV
\re{noi}. As a consequence, it is the  quantity
$\overline{m}_3^{\,2}(T)=m_3^{\,2}+\Delta m_3^{\,2}(T)$ that really controls
the existence
of membranes in the thermal bath.
Since $\Delta m_3^{\,2}(T)$ may be sizeble and negative \re{noi},
$\overline{m}_3^{\,2}(T)$ may be  small and membranes may exist
during the electroweak phase transition for zero-temperature values of $M_A$
that are in agreement with present experimental bounds \re{inprep}.

At temperatures above the critical phase-transition temperature, the thermal
fluctuations may spontaneously and abundantly produce membrane-like
configurations. A naive estimate of the number density of membranes of size $R$
produced at temperature $T$ due to thermal fluctuations is $n(R,T)\sim T^3{\rm
e}^{-F/T}$, where $F$ is the free energy of the membrane of size $R$. An
educated guess is $F\sim \eta R^2$ where $\eta$ is the energy per unit area.
The membranes have
$\eta\sim 2 M_A(T) v^2(T)$,
where $M_A^{\,2}(T)=\overline{m}_3^{\,2}(T)\sin2\beta/2$, and a typical size
$R\sim
M_A^{-1}(T)$ so that the associated free energy is given by $F\sim 2
v^2(T)/M_A(T)$. The thermal nucleation rate $\Gamma$ at which they are formed
is of the order
of $T{\rm e}^{-F/T}$ and  is much higher than the expansion rate of the
Universe
$H\sim T^2/M_{P\ell}$  $(M_{P\ell}=1.2\times 10^{19}$ GeV being the Planck
mass) as long as $F/T<{\rm ln}(M_{P\ell}/T)$. To get a feel for the numbers: At
$T\sim$ 100 GeV, $F/T$ should be smaller than 40 or so.

Membranes are expected to be  produced in  great abundance by thermal
fluctuations.  They decay just
as fast, however, since their lifetime $\tau$  is determined by interactions
with the surrounding plasma: $\tau\sim T^{-1}$.

At the electroweak phase transition, taking place at temperatures of the order
of 100 GeV, the  Standard Model gauge group
$SU(2)_{\rm L}\otimes U(1)_Y$
breaks  and the scalar fields acquire vacuum expectation values
$\langle\phi_{1,2}(T)\rangle$. The transition may occur
via nucleation of critical bubbles of radius $R_c$ (first-order phase
transition) or by an anomalous growth of  initial thermal fluctuations
in the  unstable modes  (second-order phase transition).
Membranes may still be
thermally nucleated during this epoch, and their presence  can affect the fate
of the baryon
asymmetry
produced in the transition itself if it is of the first order \re{bau1}.
In any scenario where the baryon asymmetry is generated
during a first-order electroweak phase transition, the asymmetry is
produced in the vicinity of critical bubble walls, and a strong constraint on
the ratio between the vacuum expectation value of the Higgs field
inside the bubble and the temperature must be imposed,
$\langle\phi(T)\rangle/T>1$, where in our case
$\phi(T)=\sqrt{\phi_1^{\,2}(T)+\phi_2^{\,2}(T)}$ \re{bau1}. This bound is
necessary for
the just created baryon asymmetry to survive the anomalous baryon-violating
 interactions inside the critical bubble, and may be
translated into a severe upper bound on the physical mass $M_h$ of the
scalar Higgs particle. Combining this bound with the LEP
constraint  rules out the possibility of electroweak
baryogenesis in the Standard Model of electroweak interactions,
but leaves  room for electroweak baryogenesis in
the Minimal Supersymmetric extension of the Standard Model
[\ref{toni},\ref{ew5}].

Since the rate of anomalous baryon-number-violating
processes scales like ${\rm exp}(-\langle\phi\rangle /T)$, it is clear
that even a small change in the vacuum expectation value of the Higgs
scalar field from its equilibrium value may be crucial for electroweak
baryogenesis considerations.  Because  $R_c\gg M_A^{-1}(T)$, membranes may be
thermally produced in large numbers  inside the critical bubbles
and eventually decay. Since the  vacuum expectation value
$\langle\phi(T)\rangle$ is reduced  inside the membranes with respect to  the
value in the
exterior of the membranes (see Fig.~2), baryon-number-violating processes may
be activated
in the membranes, causing a reduction of any preexisting baryon asymmetry. The
spontaneous violation of CP inside the membranes may also play a significant
role in this respect.
These and other considerations are now under investigation \re{inprep}.

\subsection*{ACKNOWLEDGMENTS}

We thank M. Carena and C. Wagner for helpful discussions. We also thank
the Nordita/Uppsala Astroparticle Programme workshop, where these ideas first
took form, for kind hospitality and support.
The work of A.R. and O.T. is supported in part by the DOE and NASA
under Grant NAG5--2788. O.T. is also supported
by the Swedish Natural Science Research Council (NFR).

%%%%%%%%%%%%%%%%%%%%%%%%%%%%%%%%%%%%%%%%%%%%%
%%%%%%%%%%%%%%%%%%%%%%%%%%%%%%%%%%%%%%%%%%%%%

\def\NPB#1#2#3{Nucl. Phys. {\bf B#1} (19#2) #3}
\def\PLB#1#2#3{Phys. Lett. {\bf B#1} (19#2) #3}
\def\PLBold#1#2#3{Phys. Lett. {\bf#1B} (19#2) #3}
\def\PRD#1#2#3{Phys. Rev. {\bf D#1} (19#2) #3}
\def\PRL#1#2#3{Phys. Rev. Lett. {\bf#1} (19#2) #3}
\def\PRT#1#2#3{Phys. Rep. {\bf#1} (19#2) #3}
\def\ARAA#1#2#3{Ann. Rev. Astron. Astrophys. {\bf#1} (19#2) #3}
\def\ARNP#1#2#3{Ann. Rev. Nucl. Part. Sci. {\bf#1} (19#2) #3}
\def\MPL#1#2#3{Mod. Phys. Lett. {\bf #1} (19#2) #3}
\def\ZPC#1#2#3{Zeit. f\"ur Physik {\bf C#1} (19#2) #3}
\def\APJ#1#2#3{Ap. J. {\bf #1} (19#2) #3}
\def\AP#1#2#3{{Ann. Phys. } {\bf #1} (19#2) #3}
\def\RMP#1#2#3{{Rev. Mod. Phys. } {\bf #1} (19#2) #3}
\def\CMP#1#2#3{{Comm. Math. Phys. } {\bf #1} (19#2) #3}

\newpage
\frenchspacing
%%%%%%%%%%%%%%%%%%%%%%%%%%%%%%%%%%%%%%%%%%%%%
%%%%%%%%%%%%%%%%%%%%%%%%%%%%%%%%%%%%%%%%%%%%%
%\begin{picture}(400,50)(0,0)
%\put (50,0){\line(350,0){300}}
%\end{picture}

%\vspace{0.25in}

\def\labelenumi{[\theenumi]}
\begin{enumerate}
\item\label{review} For a review, see, H.P. Nilles, Phys. Rep.
{\bf 110} (1984) 1; H.E. Haber and G.L. Kane, Phys. Rep. {\bf 117} (1985) 75;
A. Chamseddine, R. Arnowitt and P. Nath, {\it Applied N=1 Supergravity},
World Scientific, Singapore (1984).

\item\label{noi} D. Comelli and M. Pietroni,  Phys. Lett. {\bf B306}
(1993)
67; D. Comelli, M. Pietroni and A. Riotto, Nucl. Phys. {\bf
B412}
(1994) 441;  Phys. Rev. {\bf D50} (1994) 7703;
Phys. Lett. {\bf 343} (1995) 207;
J.R. Espinosa, J.M. Moreno, M. Quir\'{o}s,  Phys. Lett. {\bf
B319} (1993) 505.

\item\label{reviews} For a  review, see:
A.G. Cohen, D.B. Kaplan and A.E. Nelson,
Annu. Rev. Nucl. Part. Sci. {\bf 43} (1993) 27;

\item\label{sp} V.A. Kuzmin, V.A. Rubakov and M.E. Shaposhnikov,
Phys. Lett. {\bf B155} (1985) 36.

\item\label{bau} P. Huet and A.E. Nelson,
Phys. Rev. {\bf 53} (1996) 4578

\item\label{toni} M. Carena, M. Quiros,  A. Riotto, I. Vilja
and C.E.M Wagner, Fermilab--PUB--96/271-A, CERN-TH-96-242,
hep-ph/9702409, submitted to Nucl. Phys. {\bf B}.

\item\label{walls} Ya.B. Zeldovich, I.Yu. Kobzarev and L.B. Okun, Sov. Phys.
JETP {\bf 40} (1975) 1; T.W.B. Kibble, J. Phys. {\bf A9} (1976) 1387; T.W.B.
Kibble, Phys. Rep. {\bf 67} (1980) 183;
A. Vilenkin, Phys. Rep. {\bf 121} (1985) 263.

\item\label{bt} C. Bachas and T.N. Tomaras, Phys. Rev. Lett. {\bf 76} (1996)
356.

\item\label{quiros} For a recent review, see M. Quiros, {\it Lectures given at
24th International Meeting on Fundamental Physics: From the
Tevatron to the LHC\/}, Playa de Gandia, Valencia, Spain, 22-26 Apr 1996,
hep-ph/9609392

\item\label{Carena} M. Carena, J.R. Espinosa, M. Quiros and C.E.M. Wagner,
Phys. Lett. B {\bf 355} (1995) 209.

\item\label{numrec} W.H. Press, S.A. Teulkovsky,
W.T. Vetterling and B.P. Flannery, {\em Numerical Recipes in Fortran\/},
Second Edition, Chapter 17.3 (Cambridge Univ. Press, 1992).

\item\label{Eggleton}
P.P. Eggleton, Mon. Not. Roy. Astr. Soc. {\bf 151} (1971) 351.

\item\label{gp}
H. Georgi and A. Pais, Phys.~Rev.~{\bf D10} (1974) 1246.

\item\label{aleph} Talk given by G. Cowan at CERN on Feb. 25, 1997 on physics
results from the LEP run at 172 GeV. URL:
http://alephwww.cern.ch/ALPUB/seminar/Cowan-172-jam/cowan.html

\item\label{inprep} A. Riotto and O. T\"{o}rnkvist, in preparation.

\item\label{bau1}  For a recent review, see for instance, V. A. Rubakov and
M. E. Shaposhnikov, Usp. Fiz. Nauk. {\bf 166}, 493 (1996).

\item\label{ew5} J.R. Espinosa, Nucl. Phys. {\bf B475} (1996) 273;
B. de Carlos and J.R. Espinosa, SUSX-TH-97-005 preprint, hep-ph/9703212.

\end{enumerate}
\newpage
%\noindent
%{\bf\large Figure Captions}
%\vspace{1cm}

%\noindent
%{\bf Fig. 1}: Region of parameter space
%$(M_A,\tanb)$ where membrane solutions exist, for two different values of
%the supersymmetry-breaking mass $\ms$. Solutions exist below and to the
%left of  the curves.

%\vspace*{24pt}
%\noindent
%{\bf Fig. 2}:  The membrane solution for $M_A=50$ GeV, $\tanb=0.5$ and
%$\ms=5$ TeV.

%\newpage

\end{document}